\documentclass[showpacs,amsmath,preprint,showkeys,pra]{revtex4-1}

\usepackage{dcolumn}    
\usepackage{bm}         
\usepackage{ifpdf}
\usepackage{amssymb,lineno,amsfonts}
\usepackage{graphicx}   
\usepackage{bbm}
\usepackage{mathrsfs}
\usepackage{upgreek}
\usepackage{epstopdf}
\usepackage{setspace}
\usepackage{hyperref}
\usepackage[matrix,frame,arrow]{xypic}
\usepackage{float}
\usepackage{soul}
\usepackage{tikz}
\usepackage{natbib}
\usepackage{color} 
\usepackage{subfigure}
\usepackage{tikz}
\usepackage[]{qcircuit}

\definecolor{med-blue}{RGB}{25,25,112} 
\hypersetup{colorlinks, linkcolor={blue},citecolor={blue}, urlcolor={black}}
\newcommand{\ket}[1]{\vert{#1}\rangle}

\newcommand{\outpr}[2]{\vert{#1}\rangle\langle{#2}\vert}
\newcommand{\inpr}[2]{\langle{#1}\vert{#2}\rangle}

\newcommand{\proj}[1]{\outpr{#1}{#1}}

\newcommand{\tr}{\mathrm{Tr}}

\usepackage{bm}         
\usepackage{amssymb,lineno,amsfonts}
\usepackage{graphicx}   
\usepackage{bbm}
\usepackage{mathrsfs}
\usepackage{upgreek}
\usepackage{graphicx}
\usepackage{amsmath}
\usepackage{multirow}
\usepackage{diagbox}

\definecolor{med-blue}{RGB}{25,25,112} 
\hypersetup{colorlinks, linkcolor={blue},citecolor={blue}, urlcolor={black}}

\newcommand{\commute}[2]{\left[ {#1} , {#2} \right]}

\begin{document}

\title[Quantum Optimal Control]{Quantum Optimal Control: \\  Practical Aspects and Diverse Methods}
\author{T S Mahesh}
\email{mahesh.ts@iiserpune.ac.in}
\author{Priya Batra}
\email{priya.batra@students.iiserpune.ac.in}
\author{M. Harshanth Ram}
\email{m.harshanthram@students.iiserpune.ac.in}
\affiliation{Department of Physics and NMR Research Center,\\
	Indian Institute of Science Education and Research, Pune 411008, India}


\begin{abstract}
Quantum controls realize the  unitary or nonunitary operations employed in quantum computers, quantum simulators, quantum communications, and other quantum information devices. They implement the desired quantum dynamics  with the help of electric, magnetic, or electromagnetic control fields.  Quantum optimal control (QOC) deals with designing an optimal control field modulation that most precisely implements a desired quantum operation with minimum   energy consumption and maximum robustness against hardware imperfections as well as external noise.  Over the last two decades, numerous QOC methods have been proposed. They include asymptotic methods, direct search,  gradient methods, variational methods, machine learning methods, etc. In this review, we shall introduce the basic ideas of QOC, discuss practical challenges, and then take an overview of the diverse QOC methods.  
\end{abstract}

\keywords{quantum optimal control, quantum gate, state preparation, quantum dynamics}

\maketitle

\section{Introduction \label{sec:intro}}
The quest for control over quantum dynamics has a long history of several decades  (eg. \cite{PhysRevA.37.4950,KOSLOFF1989201,doi:10.1126/science.279.5358.1875,doi:10.1126/science.288.5467.824}).
Although quantum control methods have been employed in several fields from chemical kinetics to imaging, here we focus on the methods employed for quantum information related areas.
Thanks to the ongoing revolution of quantum technologies, quantum control methods have received a major impetus in recent years \cite{dowling2003quantum}. 

In this review, we mostly use examples from nuclear magnetic resonance (NMR), although the basic principles are applicable in other architectures as well.  
The early methods of quantum control for NMR quantum information tasks involved bandwidth selective fields, called shaped pulses \cite{cavanagh1996protein}.  Several quantum information tasks were realized using such band-width selective pulses \cite{dorai2000quantum}.  Although the shaped pulses are convenient to realize, they suffered from long pulse-durations as well as undesired evolutions.  
The need for precise control over quantum dynamics led to the development of advanced control methods based on optimal control theory.
Classical optimal control theory uses Pontryagin's maximum principle to design the best controls for a given task within certain constraints
\cite{kopp1962pontryagin,pontryagin1987mathematical,kirk2004optimal,boscain2021introduction,PRXQuantum.2.030203}.
Extending the control problem to the framework of quantum theory yields quantum optimal control (QOC) theory 
\cite{werschnik2007quantum,cong2014control,glaser2015training,d2021introduction}. 

This review is an attempt to provide a broad overview of various methods, without getting into finer details.  While the review is surely non exhaustive, we intend to capture several strategies and convey their gist, which may benefit a reader novice in the field. The review is arranged as follows. The next section outlines the quantum control problem formulated for quantum processors.  Section 3  discusses various practical aspects which must be taken into consideration while designing the quantum control.  Section 4 contains the review of a diverse collection of QOC methods.  
Finally, section 5 provides a summary and outlook. 

\section{Quantum Control Problem \label{sec:qcprob}}

\subsection{The system, control fields, and environment \label{sec:qproc}}
We consider an overall Hamiltonian of the form
\begin{align}
H(t) = H^S + H^\Omega(t) + H^E(t).
\label{eq:fullH}
\end{align}
Here, $H^S$ is the static \textit{system Hamiltonian}.  In quantum information processing (QIP), $H^S$ includes both self Hamiltonians of various qubits as well as their mutual interactions.  
The time-dependent \textit{control Hamiltonian} $H^\Omega(t)$ is used to realize the desired dynamics.  
The remaining undesirable interactions are collected in the effective   \textit{environmental Hamiltonian} $H^E(t)$, which may not be completely known.
We ignore this term unless we want to see its effects or design control sequences that minimize them.
The instantaneous state $\rho(t)$ of the processor satisfies the Liouville–von Neumann equation
\begin{align}
\dot{\rho}(t) = -i\commute{H(t)}{\rho(t)},
\label{eq:vn}
\end{align}
whose solution is of the form
\begin{align}
\rho(t) &= U(0,t) \rho(0) U^\dagger(0,t),~~\mbox{with unitary operator}
\nonumber \\
U(0,t) &= D \exp \left[
-i\int_{t'=0}^{t} dt' H(t') 
\right].
\label{eq:uoft}
\end{align}
Here $D$ is the Dyson's time-ordering operator.  Throughout the review, we have set $\hbar=1$.

\subsection{Control Sequence via Discretized Control Field \label{sec:cfield}}
\begin{figure}
	\centering
	\includegraphics[trim=1.5cm 0cm 3cm 0cm,width=10cm,clip=]{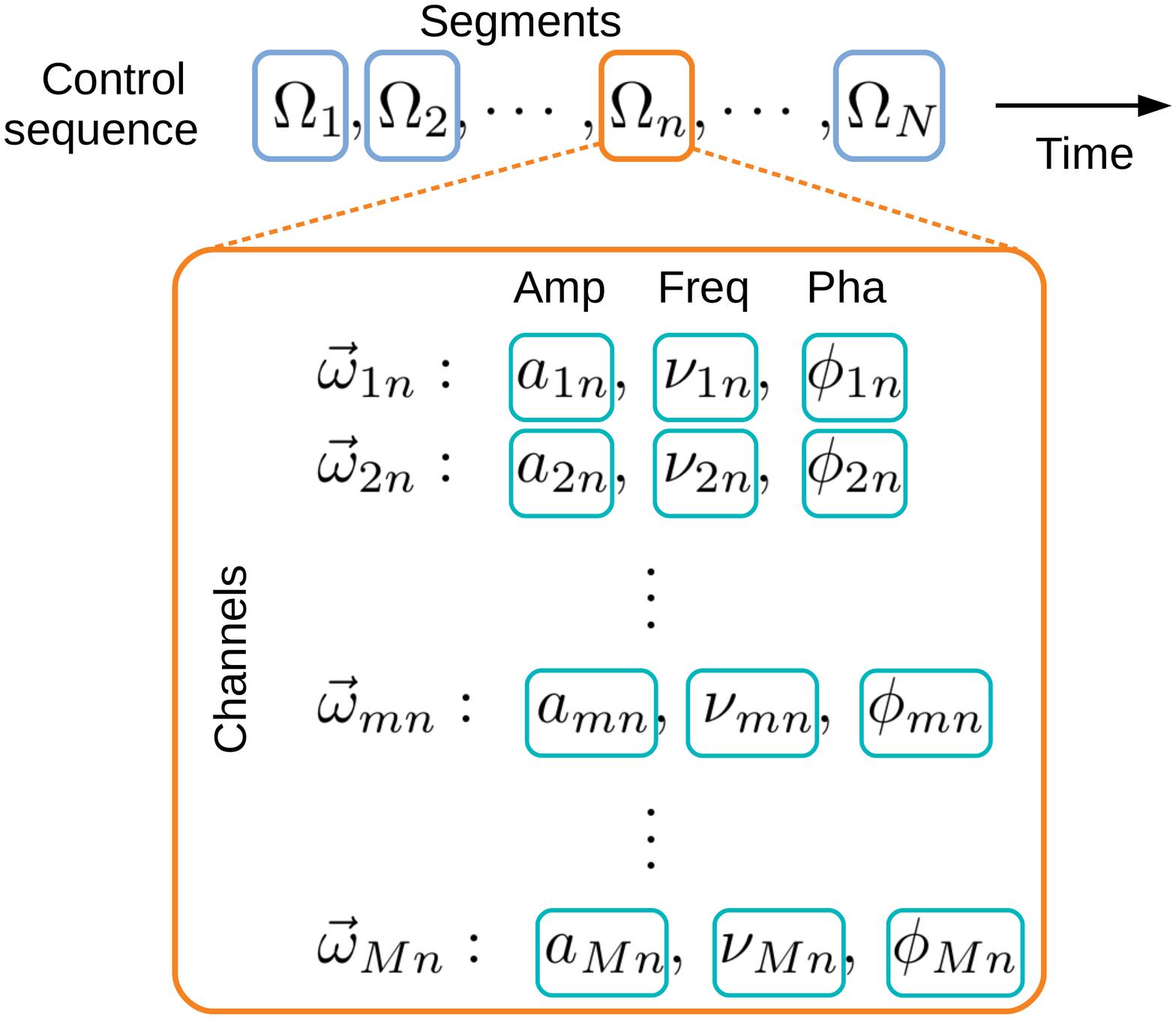}
	\caption{An example structure of the control sequence.}
	\label{fig:cseq}
\end{figure}

\begin{table*}
\begin{center}
\begin{minipage}{\textwidth}
\caption{Examples of control fields used in various quantum processor architectures. \label{tab:cfield}}
\begin{tabular}{p{2.5cm}|p{4cm}|p{3cm}|p{4cm}|p{2cm}}
\hline
Quantum Processor & 
System interactions & 
Control field(s)  &
Additional control(s) & 
Eg. references
\\ \hline \hline
Trapped atoms/ions &
Electric or 
Magnetic Dipole interactions /
Coulomb interaction between ions
  & 
Optical / Microwave 
  &
Electric / Magnetic fields
  &
\cite{schafer2020tools,haffner2008quantum}
\\ \hline
Quantum dots & 
Zeeman interaction, 
spin-spin interactions between electrons in adjacent dots
&
Microwave
&
Electric / Magnetic fields
&
\cite{veldhorst2015two,watson2018programmable}
\\ \hline
Super-conducting circuits & 
Qubit capacitance, qubit-resonator coupling capacitance, qubit-qubit coupling capacitance &
Microwave &
Magnetic field & 
\cite{krantz2019quantum}
\\ \hline
NV centers in diamond & 
Zero-field splitting, hyperfine interaction, nuclear quadrupole interaction &
Microwave for electronic spin and RF for nuclear spin &
Laser, Magnetic field, Electric field & \cite{bucher2019quantum}
\\
\hline
NMR & 
Zeeman interaction, spin-spin interaction &
RF & 
Magnetic field gradients & \cite{fortunato2002design,khaneja2005optimal,levitt2013spin} \\
\hline
\end{tabular}
\end{minipage}
\end{center}
\end{table*}
The control field may be an electric field (eg. \cite{bayer2001coupling}), or magnetic field (eg. \cite{tiwari2021universal}), or electromagnetic field (eg. \cite{jelezko2004observation}), or a combination of these (eg. \cite{pfender2017protecting}).  Table \ref{tab:cfield} lists control fields in various QIP architectures.
To bypass the complicated integral in Eq. \ref{eq:uoft} and facilitate the numerical evaluation of the propagator, we shall discretize the entire control field $\Omega(t)$ of duration $T$ into $N$ piecewise constant segments $\{\Omega_1,\Omega_2,\cdots,\Omega_n,\cdots,\Omega_N\}$, collectively termed as the control sequence $\Omega$.  Discretization by itself need not be an approximation, since the modern digital 
signal generators can produce the discretized control sequence directly (more on this in Sec. \ref{sec:nonidealcf}).
In general, each control segment $\Omega_n$ is a multi-component array with $M$ distinct channels each with one or more distinct controls (see Fig. \ref{fig:cseq}).
For example, a typical NMR spectrometer is equipped with two to five radio-frequency (RF) channels that allow simultaneous control of different nuclear spin isotopes \cite{levitt2013spin}.  In the case of NV centers, one channel controls the electronic spin and the remaining channels control nuclear spin isotopes $^{15}$N and $^{13}$C \cite{doherty2013nitrogen}. Typically, each channel has several control parameters such as amplitude, frequency, and initial phase.  Collectively, we shall represent the controls of segment $n\in[1,N]$ in channel $m\in[1,M]$ as a control vector $\vec{\omega}_{mn}$.  Amplitude is the most common control parameter since it is easier to execute in theory as well as in the experiment.  In the following, for the case of a single control parameter, we shall drop the vector symbol and denote it by $\omega_{mn}$ (see Fig. \ref{fig:cseq}).  In general, the control Hamiltonian for the $n$th segment is of the form
\begin{align}
H^{\Omega_n} = \sum_{m=1}^M  {\cal H}_{m}(\vec{\omega}_{mn}),
\label{eq:homegan}
\end{align}
with the $m$th channel control Hamiltonian ${\cal H}_m(\vec{\omega}_{mn})$.
The discretized Hamiltonian for the $n$th segment of duration $\tau_n$ and the corresponding segment propagator are
\begin{align}
H_n = H^S + H^{\Omega_n},~~
U_n &= e^{-i H_n \tau_n}.
\label{eq:un}
\end{align}
Thus the overall propagator is simply the ordered product 
\begin{align}
U(0,T) = U_{1:N} &= U_N U_{N-1} \cdots U_n \cdots U_2 U_1.
\end{align}
If segments are all of the equal duration, $\tau_n = \tau = T/N$.  

The control need not be entirely unitary. For example, we can insert projective measurements  or controlled dephasing (twirl operations \cite{anwar2005practical,bhole2016steering}) in between unitary segments, and realize a net nonunitary propagator. If $\rho_{n-1}$ represents the state after $n-1$ segments, 
\begin{align}
\rho_{n} = 
\hat{{\cal E}}_{n} \left[U_n\rho_{n-1} U_n^\dagger\right],
\label{eq:nonunievol}
\end{align}
where $\hat{{\cal E}}_{n}$ is the superoperator implementing the $n$th non-unitary operation.  We now discuss two types of quantum control problems.

\subsection{Quantum Gate Synthesis Problem \label{sec:qgates}}
Given a target unitary operator $U_F$ corresponding to a quantum gate, the goal here is to find a control sequence $\Omega$ that maximizes the gate fidelity \cite{nielsen2002quantum}
\begin{align}
F_G(U_F,U_{1:N}) = \left\vert\frac{\inpr{
U_F}{U_{1:N}}}{\inpr{U_F}{U_F}}\right\vert^2.
\label{eq:fu}
\end{align}
Here and throughout the rest of the article we have used 
\begin{align}
\inpr{A}{B} = \tr\left(
A^\dagger B
\right)
\end{align}
to denote the overlap between the operators $A$ and $B$. 
The control sequence so obtained will be applicable independent of the initial state.

\subsection{State to State Transfer or State Preparation Problem
\label{sec:s2s}}
Here the control sequence $\Omega$ needs to transfer a specific initial state $\rho_{I}$ to a specific final state $\rho_{F}$.  If the two states are of the same purity, we seek a unitary sequence  $U_{1:N}$ that prepares the state
\begin{align}
\rho_N &= U_{1:N} \rho_I U_{1:N}^\dagger
\end{align}
as close to $\rho_F$ as possible.
Examples include adiabatic inversion/excitation
\cite{cavanagh1996protein,tannus1997adiabatic}, preparing singlet-triplet order \cite{manu2012singlet,khurana2017bang}, etc.

Often, the initial and the final desired states need not have  the same purity.   For instance, preparing pseudopure state in NMR from thermal state \cite{cory1997ensemble,suter2008spins,bhole2016steering}. In such a case we seek a nonunitary sequence as described in Eq. \ref{eq:nonunievol}.

In state preparation, the control sequence $\Omega$ is obtained by maximizing one of the following measures.
\begin{itemize}
\item[(i)] Uhlmann state-fidelity (eg. \cite{wu2012control}):
\begin{align}
F_S(\rho_F,\rho_{N}) = \left( 
\tr\sqrt{\sqrt{\rho_F} \rho_{N} \sqrt{\rho_F}} \right)^2.
\label{eq:fs}
\end{align}
\item[(ii)] Trace fidelity (eg. \cite{khaneja2005optimal}):
\begin{align}
F_T(\rho_F,\rho_{N}) = \inpr{\rho_F}{\rho_{N}}.
\label{eq:ft}
\end{align}
\item[(iii)] For traceless parts of the density matrices, it is particularly convenient to use the following measures \cite{fortunato2002design}:
\begin{itemize}
\item Correlation:
\begin{align}
F_C(\rho_F,\rho_{N}) &= \frac{\inpr{\rho_F}{\rho_{N}}}
{\sqrt{\tr(\rho_F^2)} \cdot \sqrt{\tr(\rho_{N}^2)}}.
\nonumber \\
\end{align}
\item
Attenuated correlation:
\begin{align}
F_A(\rho_I,\rho_F,\rho_{N}) &= \frac{\inpr{\rho_F}{\rho_{N}}}
{\sqrt{\tr(\rho_I^2)} \cdot \sqrt{\tr(\rho_{N}^2)}},
\label{eq:fca}
\end{align}
which is also sensitive to the change in the purity of $\rho_F$ w.r.t. the initial state $\rho_I$, in addition to the overlap between $\rho_N$ and $\rho_F$.
\end{itemize}
\end{itemize}

While it is relatively easier to generate a state-specific control sequence compared to the the general quantum gate discussed in Sec. \ref{sec:qgates}, the state to state control sequence generated for a particular pair of initial and final states is not applicable for other states.

\section{Practical Aspects \label{sec:pract}}
\subsection{Control field limitations
\label{sec:nonidealcf}}
\subsubsection{Distribution of a control parameter
\label{sec:inhomo}}
\begin{figure}
	\centering
	\includegraphics[trim=0cm 0cm 0cm 0cm,width=9cm,clip=]{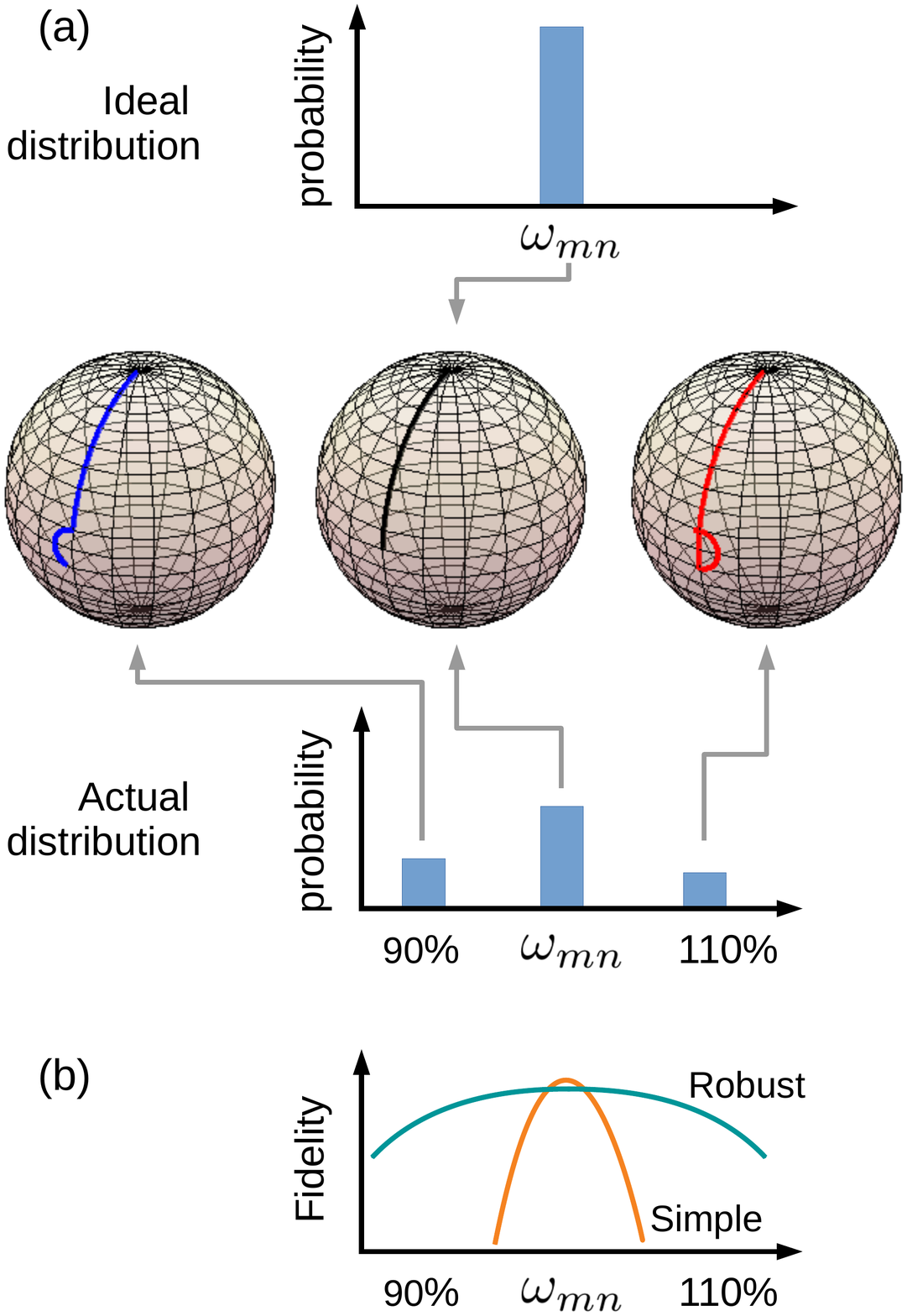}
	\caption{(a) Illustrating ideal (top) and typical (bottom) spatial distribution of RF amplitude about a nominal value of $\omega_{mn}$.  Middle row: the Bloch-sphere trajectories on applying  Hadamard gate $H = R_x(\pi) R_y(\pi/2)$ on $\ket{0}$ with 10\% weaker (left), nominal (center), and 10\% stronger (right) field-components.  Note that, only for the nominal component, the qubit reaches the desired state $(\ket{0}+\ket{1})/\sqrt{2}$. 
	(b) The fidelity profiles of simple and robust control sequences.}
	\label{fig:traj}
\end{figure}

Suppose a control segment with a nominal amplitude value $\omega_{mn}$ actually has a distribution over the range
$[\omega_{mn}^\mathrm{min},\omega_{mn}^\mathrm{max}]$.
We may discretize the distribution into $L$ bins, and assign probability $p_l$ for the $l$th bin (see Fig. \ref{fig:traj}). 
It is important to have a quantitative understanding of the distribution.  For example, Pravia et al. \cite{pravia2003robust} described the experimental estimation of RF distribution via Torrey oscillation.
A microscopic system, like an NV center, may sample one particular bin of the distribution, while a macroscopic object, such as an NMR spin ensemble, experiences the entire distribution  \cite{cavanagh1996protein}.  
To realize a control sequence ${\Omega}$ that is robust against the distribution, we optimize the weighted average by defining the performance function as \cite{fortunato2002design,khaneja2005optimal},
\begin{align}
\Phi(\Omega) = \bar{F} = \sum_{l=1}^{L} p_{l} F_l,
\label{eq:fbar}
\end{align}
where $F_l$ is the fidelity of the $l$th bin (see Fig. \ref{fig:traj}(b)). 

\subsubsection{Power restriction \label{sec:powerbw}}
The hardware limitation imposes a maximum nominal value of $\omega_m^{\mathrm{MAX}}$ for each control parameter $\omega_{mn}$.  To restrict the control sequence ${\Omega}$ within the allowed range, we maximize the improved performance function
\begin{align}
\Phi(\Omega) = \bar{F} -{\cal P}(\{\omega_m^{\mathrm{MAX}}\}),
\label{eq:phi}
\end{align}
where the penalty function ${\cal P}$ is normally zero, but rapidly grows as any control parameter reaches close to or exceeds the limits  \cite{fortunato2002design,fletcher1983penalty}. 

\subsubsection{Frequency bandwidth and nonlinear response  \label{sec:freqbw}}
Electronic circuits producing a control field are designed for a certain maximum frequency bandwidth, beyond which the signal profiles are distorted.  This limits how fast the control parameters can be changed from one segment to another.  To this end, we may seek a control sequence in terms of low-bandwidth and smoothly varying basis functions, such as Slepian functions \cite{lucarelli2018quantum,feng2018gradient}. A related issue is the finite raise-time \& fall-time, as well as other nonlinearities in a practical pulse.  Boulant et al. \cite{boulant2003experimental} used an interesting approach  that involved reading the actual nonlinear pulse profile as seen by a spy detector coil. This enabled them to model the nonlinear transfer function, find its inverse, and iteratively obtain a compensated pulse.

\subsection{Noise and decoherence
\label{sec:noise}}
As mentioned in Eq. \ref{eq:fullH}, a quantum system being controlled is often prone to certain unavoidable environmental fields or interactions, leading to decoherence. The challenge here is to design a protected quantum control, that is robust against these external influences. 
One often addresses this problem with a time-optimal control sequence (eg. \cite{mottonen2006high,zhang2011time,tibbetts2012exploring}), that performs the desired task before noise becomes significant.
The other powerful approach is by interlacing dynamical decoupling pulses with the control sequence to realize protected quantum gates \cite{xu2012coherence,zhang2014protected,PhysRevLett.82.2417,ram2021robust}.  If additional qubits are available, one can also encode information onto logical qubits in a decoherence free subspace (DFS)  \cite{PhysRevLett.81.2594}.  
An extensive review for controlling the open quantum system has been covered by Koch \cite{koch2016controlling}.  

\subsection{Computational resource for optimization
\label{sec:compcost}}
The computational resource needed for generating a control sequence is an important consideration in choosing the optimization method for a given control problem.  For systems with large Hilbert-space dimensions, the bottleneck is in the matrix exponentiation in Eq. \ref{eq:un}.  Some ideas are discussed below.

\subsubsection{Limiting off-diagonal operators}
Certain quantum control tasks, especially nonlocal quantum gates, require long delays (evolution under  system Hamiltonian $H^S$)  in between  control pulses. The delay propagator
$U_d(\tau)$
can be efficiently evaluated in the eigenbasis $\{\ket{s}\}$ of the system Hamiltonian $H_S$, i.e.,
\begin{align}
U_d(\tau) =  e^{-iH^S\tau} =  \sum_s e^{-is\tau} \proj{s}.
\end{align}
Therefore, a due emphasis on delays in such control sequences greatly improves the efficiency of optimization \cite{mahesh2006quantum}.

\subsubsection{Diagonalizing Control Hamiltonians}
In certain scenarios, it is possible to find a clever choice of basis operators that helps avoiding iterative matrix exponentiation \cite{bhole2017rapid,bhole2018practical}.
The trick is to use the Trotter decomposition and rewrite Eq. \ref{eq:un} as 
\begin{align}
U_n = e^{-i(H^S+H^{\Omega_n})\tau} &\approx e^{-iH^S\tau/2}
e^{-iH^{\Omega_n}\tau} e^{-iH^S\tau/2}
\nonumber \\
&~~ = U_d(\tau/2)~
e^{-iH^{\Omega_n}\tau}
~U_d(\tau/2).
\end{align}
The nontrivial central exponential is then diagonalized using  a pair of fixed unitary operators $W_1$ and $W_2$ so that
\begin{align}
e^{-iH^{\Omega_n}\tau} 
&= W_1 \left(
\sum_d e^{-id\tau_n} \proj{d}
\right) W_2,
\end{align}
using which $U_n$ can be efficiently computed \cite{bhole2018practical}.

\subsubsection{Use of fixed controls or bang-bang controls}
While designing quantum controls for  large quantum systems, often it might be sufficient to use only the maximum power pulses, also known as the bang-bang controls, separated by delays. In this case, the matrix exponentiation in Eq. \ref{eq:un} reduces to 
\begin{align}
U_n = 
\begin{cases}
& U_d(\tau_n) ~~\mbox{if delay segment, else}
\nonumber \\
& \exp\left[
-i\tau\left(
H_S + \sum_{m=1}^M \omega_m^\mathrm{max}
H_m
\right)
\right].
\end{cases}
\end{align}
Both cases are effectively one-time calculations and accordingly, the bang-bang controls can be generated efficiently even for larger quantum systems \cite{bhole2016steering,khurana2017bang}.
 
\section{Diverse Methods \label{sec:nummtd}}
\begin{figure*}[h]
	\centering
	\includegraphics[trim=2cm 0cm 4.4cm 0cm,width=17cm,clip=]{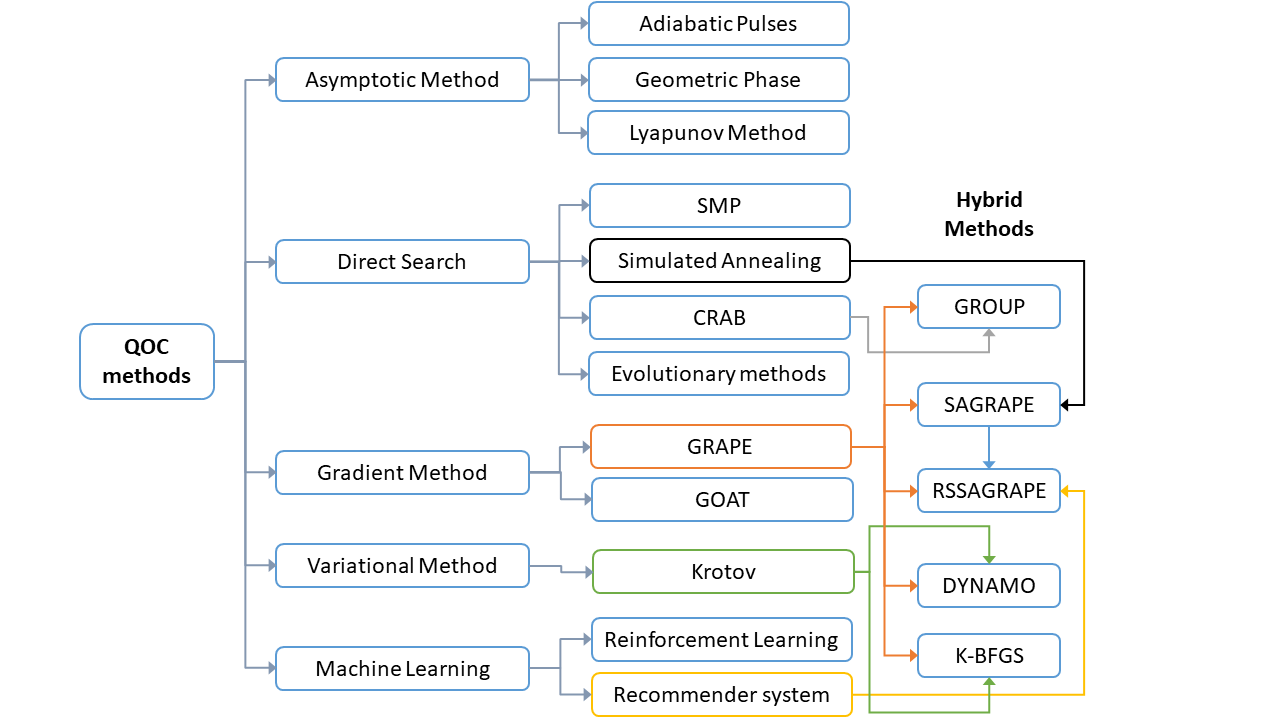}
	\caption{An overview of the diverse QOC methods discussed in this review.}
	\label{fig:dmtd}
\end{figure*}

We now discuss several of the diverse methods which have been employed for QOC (see Fig. \ref{fig:dmtd}).

\subsection{Asymptotic Evolution Methods}
The following methods assume slow variation of some parameter, and approach exact solutions in the asymptotic limit.
\subsubsection{Adiabatic pulse method}
Adiabatic pulses are considered a method of choice for state transfers in the presence of highly inhomogeneous fields \cite{cavanagh1996protein,tannus1997adiabatic,garwood2001return}.  
Consider a state transfer problem from the ground state $\ket{\psi_I}$ of an initial Hamiltonian ${\cal H}_I$ to the ground state $\ket{\psi_F}$ of the final Hamiltonian ${\cal H}_F$.  The adiabatic method involves implementing a control Hamiltonian of the form 
\begin{align}
H(t) = (1-\omega(t)){\cal H}_I + \omega(t){\cal H}_F.
\end{align}
The scalar parameter $\omega(t)$ is slowly (adiabatically) varied from $\omega(0) = 0$ to $\omega(T) = 1$, such that the system remains in an eigenstate of the instantaneous Hamiltonian $H(t)$ throughout \cite{messiah2014quantum}. 
As long as there is no level-crossing, this method leads to a robust state transfer from $\ket{\psi_I}$ to $\ket{\psi_F}$, since the field imperfections can only affect the intermediate trajectory, but not the final state.  

\begin{figure}
	\includegraphics[trim=0cm 0cm 0cm 0cm,width=10cm,clip=]{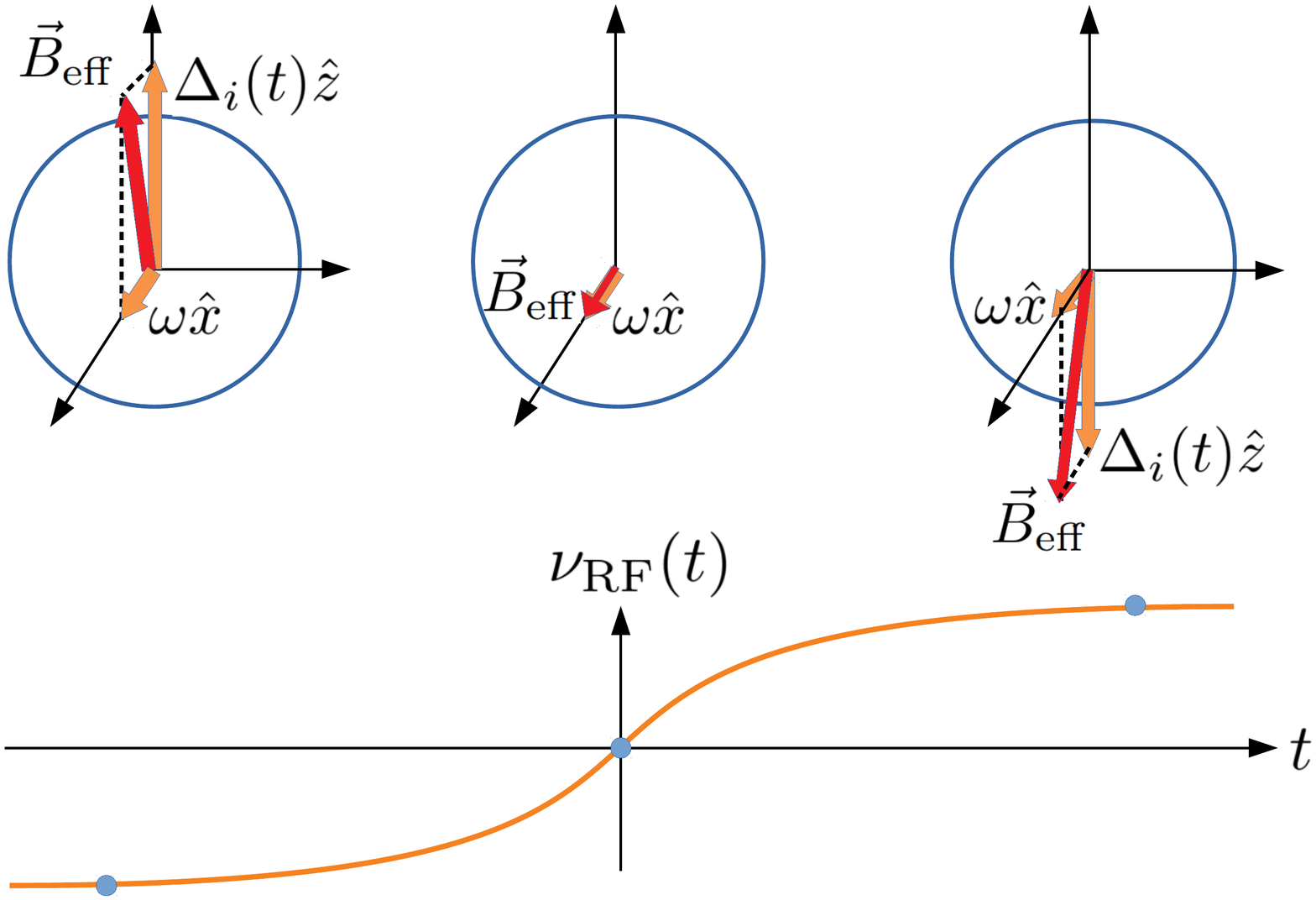}
	\caption{Different stages of adiabatic inversion.}
	\label{fig:adiabatic}
\end{figure}

Broadband inversion forms a simple, yet an intuitive example of the adiabatic method (see Fig. \ref{fig:adiabatic}). 
Imagine inverting a set of spin qubits of Larmor frequencies $\delta_i$ from $\ket{00\cdots0}$ to $\ket{11\cdots1}$ using an RF field of amplitude $\omega$, frequency $\nu_\mathrm{RF}$, and phase $\hat{x}$.
The effective Hamiltonian in the rotating frame of the RF is of the form
\begin{align}
H_\mathrm{eff}(t) = \sum_i \Delta_i(t) \sigma_z/2  +  \omega \sigma_x/2,~~\mbox{with}~~\Delta_i(t) = 2\pi\left[\delta_i-\nu_\mathrm{RF}(t)\right].
\end{align}
This corresponds to a precession of the spin qubits about an effective field $\vec{B}_{i,\mathrm{eff}} \propto  
\Delta_i(t) \cos\theta_i(t) \hat{z} + \omega\sin\theta_i(t) \hat{x}$, where $\theta_i(t) = \tan^{-1} \left( \omega/\Delta_i(t) \right)$.  Initially, one starts with a highly off-resonant pulse, i.e., $\vert\nu_\mathrm{RF}(0)\vert \gg \delta_i$ and $\vert\Delta_i(0)\vert \gg \omega$, so that $\theta_i(0) \sim 0$.  Thus, the initial state 
$\ket{00\cdots0}$ is an eigenstate of $H_\mathrm{eff}(0)$, and the qubits are locked to the effective field.
Gradually the RF frequency is swept to on-resonance ($\theta  = \pm \pi/2$) and slowly taken off-resonant far on the other side.  This way, all the qubits continue to get locked to the effective field and finally reach the state $\theta = \pi$, which corresponds to the state $\ket{11\cdots 1}$.  In the adiabatic limit, the inversion happens to all the spin qubits with different Larmor frequencies, even in the presence of inhomogeneous amplitude $\omega$.  

Adiabatic methods are routinely being applied in several fields including spectroscopy and imaging \cite{tannus1997adiabatic,garwood2001return}.
Adiabatic quantum algorithms, wherein the solution of a task is modeled as  the ground state of a Hamiltonian, are also popular \cite{albash2018adiabatic}.

\subsubsection{Geometric phase method}
If the state of a quantum system undergoes a cyclic evolution and returns to the original state \cite{berry1988geometric,anandan1992geometric}, then apart from a dynamical phase, it acquires a geometric phase, that is purely dependent on the geometric aspects of the path followed.  The dynamical phase can be effectively removed using a spin-echo type of time-reversal sequence leaving behind a pure geometric phase \cite{jones2000geometric}.  
In the adiabatic limit, the geometric phase is called the Pancharatnam-Berry phase \cite{pancharatnam1956generalized,berry1987adiabatic}.  
The geometric phase has long been observed and used in NMR \cite{suter1987berry}.  Ekert and coworkers \cite{ekert2000geometric} had proposed geometric phase quantum gates and recognized them to be robust against certain types of control errors. 
Universal quantum computing by non-abelian geometric phase gates, or holonomic gates, was proposed by Zanardi and Rasetti \cite{zanardi1999holonomic}.
Subsequently, there have been several experimental demonstrations of geometric quantum gates, including in NMR  \cite{jones2000geometric}, ion qubits
 \cite{leibfried2003experimental}, ultracold neutrons \cite{PhysRevLett.102.030404}, superconducting qubits \cite{PhysRevA.87.060303}, as well as NV centers \cite{nagata2018universal}.

\subsubsection{Lyapunov method
\label{sec:lyapunov}}
Lyapunov method was originally used for stabilizing a dynamical system into an asymptotic stable state \cite{enwiki:1033346201,isidori1995local} and is particularly useful for quantum state transfer \cite{grivopoulos2003lyapunov}.  
To explain the method, we shall consider a transfer from an arbitrary initial state $\ket{\psi_I}$ to an eigenstate $\ket{\psi_F}$ of the system Hamiltonian $H^S$ \cite{hou2012optimal,wang2014optimal}.  
We model the distance between the instantaneous state and the target state in terms of a positive function, called Lyapunov function $V$, which is  a continuously differentiable and  decreasing function, i.e., $\dot{V} \le 0$.  There can be multiple ways to set up the Lyapunov function, one of which is
\begin{align}
V(t) = \inpr{P}{\rho(t)},~~\mbox{with}~~ P = I - \proj{\psi_F},
\end{align}
and $\rho(t) = \proj{\psi(t)}$ being the instantaneous state. As the system reaches the target state $V(t)\rightarrow 0$. Now invoking the Lyapunov constraint,
\begin{align}
\dot{V}(t) &= \inpr{P}{\dot{\rho}(t)} = -i\sum_m \omega_m(t)\inpr{P}{\left[H^S +  {\cal H}_m,\rho(t) \right]} \le 0
~~~\mbox{(using Eqn. \ref{eq:vn})},
\nonumber \\
&= \sum_m \omega_m(t) v_m(t) \le 0 
~~~\mbox{with}~v_m(t) =  -i\inpr{\rho(t)}{[P,{\cal H}_m]},
\end{align}
where we have used $[P,H^S] = 0$.
 There are different ways to satisfy the above constraint.  Optimal Lyapunov solutions are discussed in Ref. \cite{hou2012optimal}.  For instance, if the control amplitudes are constrained by $\vert\omega_m (t)\vert\le \omega_m^{\mathrm{max}}$, then  one obtains the bang-bang control form \cite{hou2012optimal}
\begin{align}
\omega_m(t) = 
\begin{cases}
-\omega_m^\mathrm{max} ~&~  v_m(t) > 0,
\\
\omega_m^\mathrm{max} ~&~
 v_m(t) < 0,
\\
0 ~&~  v_m(t) = 0.
\end{cases}
\end{align}
Thus the Lyapunov method provides a systematic way to generate quantum control for state transfer from an arbitrary initial state to an eigenstate of the system Hamiltonian $H^S$.  
It is also possible to generalize the Lyapunov method for gate synthesis (eg. \cite{ghaeminezhad2018preparation}).
Wang et al \cite{wang2022quantum} described a 
Lyapunov based bang-bang quantum control scheme  for cooling  an optomechanical oscillator.
Recently, Purkayastha \cite{purkayastha2022lyapunov} extended the Lyapunov method for linear open quantum systems.

\subsection{Direct Search Methods
\label{sec:stsearch}}
Direct search methods rely on cleverly sampling the search space starting from a random initial guess.  The following methods employ a stochastic search algorithm to reach an optimal solution.
\subsubsection{Strongly Modulating Pulses (SMP)}
\begin{figure}[h]
	\centering
	\includegraphics[trim=5cm 0cm 4cm 0cm,width=9cm,clip=]{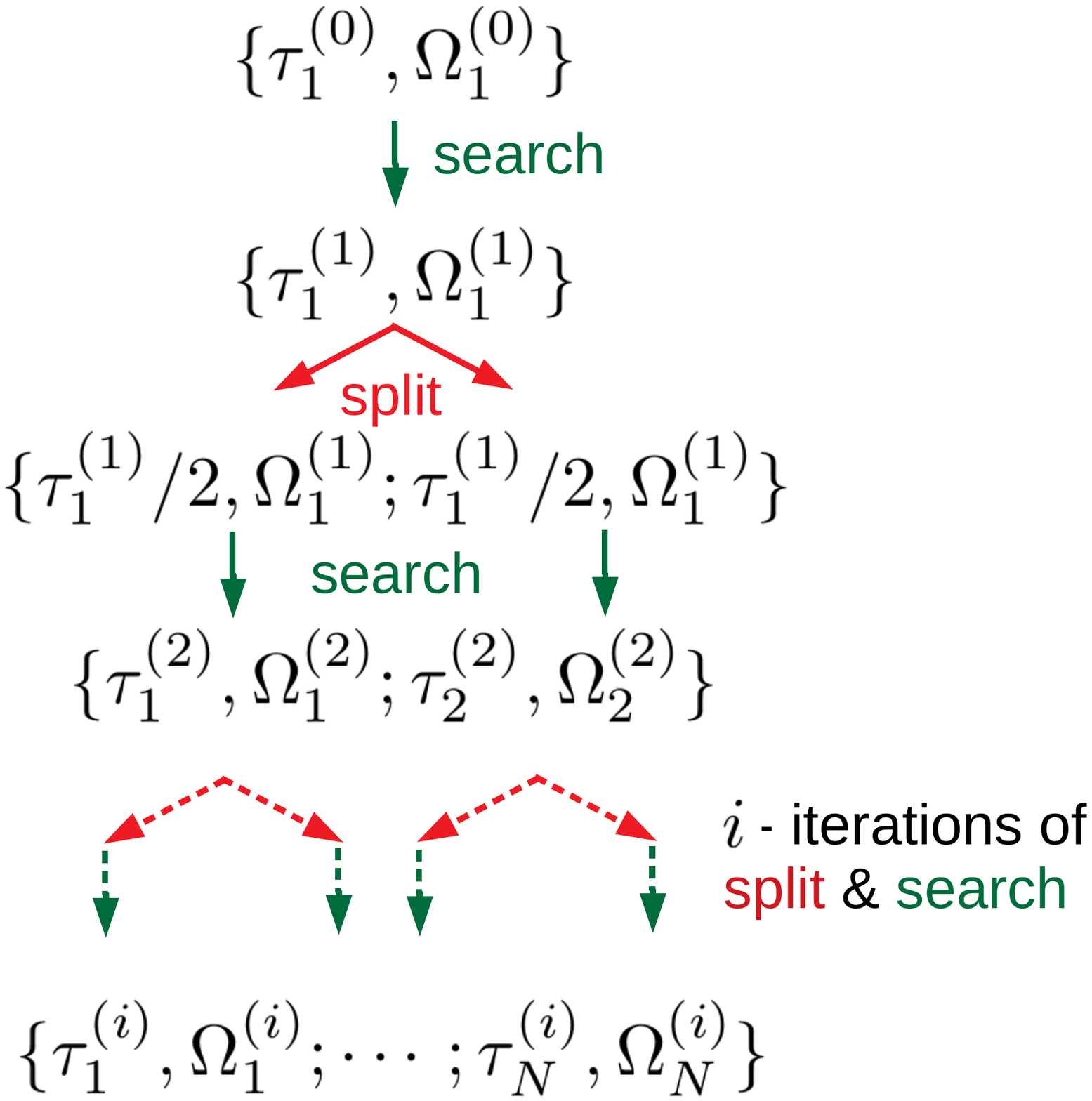}
	\caption{Split and search method used for generating SMP.}
	\label{fig:smp}
\end{figure}

Fortunato et al \cite{fortunato2002design} described generating SMPs using a stochastic search with a split and search method as illustrated in Fig. \ref{fig:smp}.  
Given a target operator $U_F$, it starts with just one random control segment $\Omega_1^{(0)}$ with a sufficiently long duration $\tau_1^{(0)}$. The Hamiltonian for the one-segment sequence is simply $H^{(0)}_1 = H^S + H^{\Omega_1^{(0)}}$, and the corresponding propagator is $U_1^{(0)} = \exp(-i H^{(0)}_1 \tau_1^{(0)})$.
By maximizing the performance function $\Phi$ (see Eq. \ref{eq:phi}) using a vector stochastic search, such as Nelder-Mead simplex \cite{nelder1965simplex}, we obtain $\{\tau^{(1)}_1,\Omega_1^{(1)}\}$.  Now we split it into two equal segments $\{\tau^{(1)}_1/2,\Omega_1^{(1)};\tau^{(1)}_1/2,\Omega_1^{(1)}\}$ and use them as the starting point for second iteration. The final solution involves a control sequence $\Omega$ wherein each segment has a specific duration, amplitude, frequency, as well as initial phase.
Although stochastic search is computationally expensive, it was successfully applied in generating controls in several QIP tasks \cite{boulant2003experimental,weinstein2004quantum,mahesh2006quantum,baugh2006solid},
including QOC of the 12-qubit NMR system \cite{negrevergne2006benchmarking}.

\subsubsection{Simulated annealing (SA)
\label{sec:simuanneal}}
\begin{figure}[h]
	\centering
	\includegraphics[trim=0cm 0cm 0cm 0cm,width=8cm,clip=]{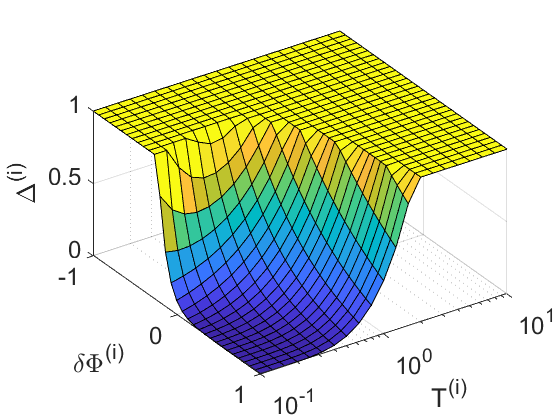}
	\caption{Threshold function $\Delta^{(i)}$ vs temperature $T^{(i)}$ and change in performance $\delta \Phi^{(i)}$.}
	\label{fig:sathreshold}
\end{figure}

SA \cite{Simul} is a  metaheuristic algorithm that attempts to reach the global optimum of the parameter space by gradually shifting from exploration mode to exploitation mode. In the $i$th iteration of threshold-based SA, the solution $\Omega^{(i)}$ with performance $\Phi(\Omega^{(i)})$   is compared against a random neighborhood point $\Omega'$ with performance $\Phi(\Omega')$.  We define
$\delta \Phi^{(i)}=\Phi(\Omega^{(i)})-\Phi(\Omega')$ and the threshold function (see Fig. \ref{fig:sathreshold})
\begin{align}
\Delta^{(i)} = \min\left[1,T^{(i)}e^{-\delta \Phi^{(i)}/T^{(i)}}\right].
\end{align}
Here, the temperature $T^{(i)}$ is an iteration-dependent parameter lowering which causes the threshold function to approach zero.
The selection rule is
\begin{align}
\Omega^{(i+1)}
&= \begin{cases}
\Omega' &  \mbox{if}~~
\delta \Phi^{(i)}  \le \Delta^{(i)}
\\
\Omega^{(i)} & \mbox{otherwise.}
\end{cases}
\end{align}
 Notice that the random solution is selected not only when it is a better solution but also when it is slightly worse than the current solution. This is the salient feature of SA that enables it to jump over local optima and reach the best neighborhood. Eventually the algorithm should promote solutions comparable to or better than the current solution.  
This is achieved by gradually reducing the temperature parameter which makes the algorithm more and more exploitative. Some recent applications of SA include solving quantum circuit transformation problem \cite{zhou2020quantum},
preparing high fidelity quantum controls  \cite{ram2021robust}, and optimizing quantum circuits for simultaneous dense protocol \cite{situ2022using}.

\subsubsection{Chopped random basis optimization (CRAB) \label{sec:CRAB}}
In QOC, often high fidelity solutions may lie in a low dimensional subspace of the larger dimensional search space \cite{lloyd2014information}. This property is exploited in algorithms like CRAB, where the control sequence  is written as a linear combination of sensibly chosen basis functions which are fewer in numbers  compared to the dimensionality of the original search space. 
CRAB was first developed for optimizing time dependent density matrix simulations \cite{PhysRevLett.106.190501}, and was later adapted for the optimization of quantum processes \cite{PhysRevA.84.022326,muller2021one}. 
Here, the control sequence 
is modeled in terms of a continuous orthogonal functional basis, such as Fourier series, Lagrange polynomials, Hermite functions, etc.  For example, in terms of  Fourier basis, the amplitude of the control sequence of duration $T$ for channel $m$ can be modeled as
\begin{align}
\omega_{m}(t) &= \bar{\omega}_m +  \sum_{k=1}^K  \alpha_{km} \cos\phi_{km}(t) + \beta_{km}\sin\phi_{km}(t),
\end{align}
Here $\bar{\omega}_m$ is the constant component,  $\alpha_{km}$, $\beta_{km}$ are the Fourier coefficients, $K$ is the total number of harmonics, and $\phi_{km} = 2\pi k t (1+r_{km})/T$ with random numbers $r_{km}\in[-0.5,0.5]$  added to improve the convergence.
Since the control sequence is not discretized, one resorts to numerical methods such as Runge-Kutta integration to calculate the propagator (Eq. \ref{eq:uoft}).  The performance function is then maximized by finding the optimal parameters for the model function.  For this,
CRAB relies on the direct search, such as Nelder-Mead simplex \cite{riaz2019optimal}.
The CRAB algorithm has been successfully applied to control of Bose-Einstein-Condensates \cite{PhysRevA.98.022119}, adiabatic population transfer of dressed spin states \cite{PhysRevA.99.063812}, quantum Szilard engine optimization \cite{PhysRevA.100.042314}, etc. 

\subsubsection{Evolutionary Methods
\label{sec:evolalgo}}
Evolutionary methods, such as genetic algorithm (GA), are inspired from biological evolution processes, such as reproduction, mutation, and survival of fittest, which leads to the emergence of a strong breed in an evolving population. In GA, the initial random population of genes goes through crossover, mutation, and selection based on a fitness function.  There have been several implementations of GA based QOC, such as  generating unitary and nonunitary QOC \cite{bhole2016steering}, for preparing singlet order in an 11-qubit register \cite{KHURANA20178}, and QOC of qutrits \cite{PhysRevA.90.032310}.
Another variant of the evolutionary method, namely differential evolution (DE),
has also been used for gate control \cite{PhysRevA.90.032310} and
QOC of  the open quantum system  \cite{ma2015differential}.

\subsection{Gradient Methods}
We now look at optimization methods that systematically move towards a local optimum by evaluating the local gradients.  
\subsubsection{Gradient Ascent Pulse Engineering (GRAPE)
\label{sec:grape}}
\begin{figure}[h]
	\centering
	\includegraphics[trim=0cm 0cm 0cm 0cm,width=12cm,clip=]{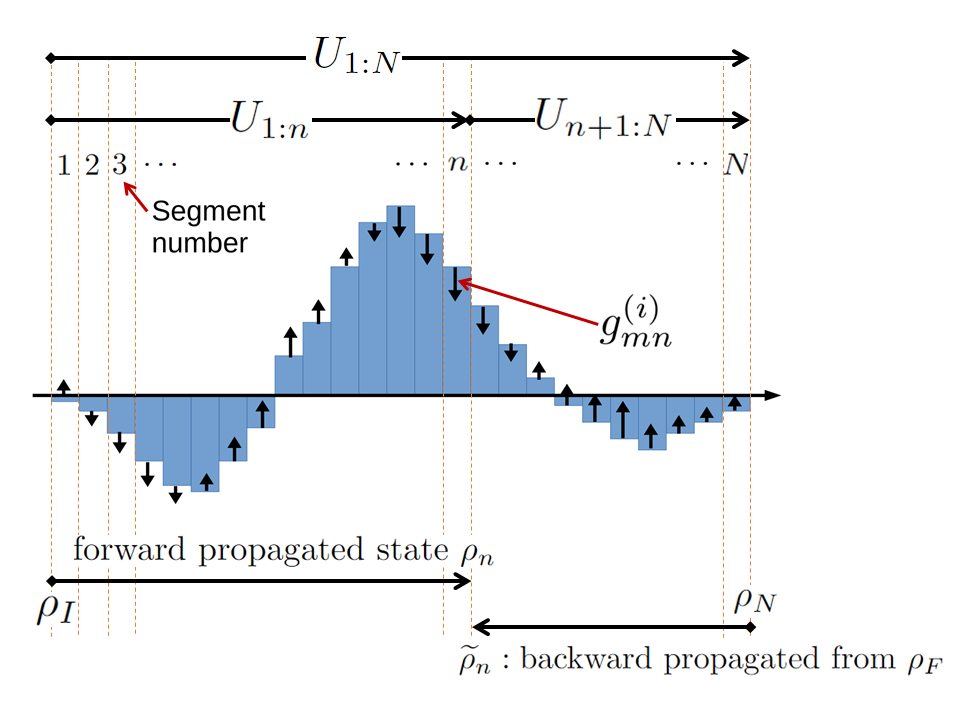}
	\caption{Illustrating the gradients $g^{(i)}_{mn}$ for amplitudes $\omega_{mn}$ in the $i$th iteration of GRAPE algorithm \cite{khaneja2005optimal}.}
	\label{fig:grape}
\end{figure}

Khaneja et al. \cite{khaneja2005optimal} 
proposed a gradient method for iteratively generating a control sequence.  Being a local search method, it starts from a random sequence ${\Omega}^{(0)}$ of $N$ segments, each of duration $\tau$, and the goal is to optimize the amplitudes $\omega_{mn}$.  The
control Hamiltonian is 
${\cal H}_m({\omega}_{mn}) = \omega_{mn} {\cal A}_m,$
where ${\cal A}_m$ is the control operator.
In the $i$th iteration, each segment is corrected according to
\begin{align}
\omega_{mn}^{(i)} &= \omega_{mn}^{(i-1)} + \epsilon \tau  ~g_{mn}^{(i)},
\end{align}
where $\epsilon$ is the step size and $g_{mn}^{(i)}$ is the gradient (see Fig. \ref{fig:grape}).
GRAPE is popular, thanks to the simple analytical expressions for the first-order gradients \cite{khaneja2005optimal}.
\begin{align}
g_{mn}^{(i)} = 
\begin{cases}
 \mbox{Im}\left\{
 \inpr{U_F}{U_{n+1:N}{\cal A}_m  U_{1:n}} \inpr{U_{1:N}}{U_F}
\right\} ~&~ \mbox{gate synthesis}
\\
 -i~ \inpr{\widetilde{\rho}_{n}}{\commute{{\cal A}_m}{\rho_{n}}}
 ~&~ \mbox{state transfer.}
 \end{cases}
 \label{eq:gmn}
\end{align}
Here forward propagated state~ $\rho_{n} = U_{1:n} \rho_I U_{1:n}^\dagger$ and 
backward propagated state $~\widetilde{\rho}_{n} =
U_{n+1:N}^\dagger \rho_F U_{n+1:N}$.

While GRAPE has been used for quantum control in multiple architectures \cite{rowland2012implementing}, there are also numerous variants of GRAPE. Lucarelli  \cite{lucarelli2018quantum} implemented  
the bandwidth limited GRAPE algorithm using Slepian basis functions.  Priya et al \cite{batra2020push} reported Push-Pull GRAPE wherein the optimization is driven by the attraction of target operators as well as the repulsion from orthogonal operators.
GRAPE for open quantum systems is discussed in \cite{boutin2017resonator}. GRAPE has also been adopted for optimal control of quantum measurement \cite{PhysRevA.90.052331}.

The convergence rate of the GRAPE algorithm is limited by the first-order approximation made in the gradient expression of Eq. \ref{eq:gmn}. Moreover, the step-size $\epsilon$ needs to be smartly chosen for a given control problem.  These limitations are overcome by the second-order method, which is computationally expensive though.  The quasi-Newton method namely,  Broyden–Fletcher–Goldfarb–Shanno (BFGS) is one of the popular second-order algorithms.  
de Fouquieres et al \cite{de2011second}  demonstrated convergence acceleration via BFGS algorithm.

\subsubsection{Gradient optimization of analytic controls (GOAT)}
While GRAPE is quite successful, the control sequences are discrete and rugged.  Like CRAB discussed earlier, one may rather prefer to express the control sequence in a smooth analytical form.  It also allows encoding the solution in a lower dimensional parameter space.  This is one of the benefits of gradient optimization of analytic controls (GOAT) \cite{PhysRevLett.120.150401}.
The model analytical function may be an educated guess or randomly chosen. For example, if one models the control sequence in terms of the superposition of $K$ Gaussian pulses,
\begin{align}
\omega_{m}(t) = 
\sum_{k=1}^K \exp
\left[
-\frac{(t-\delta_{m,k})^2}{\sigma_{m,k}^2}
\right].
\end{align}
Here $m$th channel parameters $\alpha_m = \{\delta_{m,k},\sigma_{m,k}\}$ are optimized to minimize the gate infidelity 
$f(\Omega) = 1-{\inpr{U_F}{U_\Omega}}/{\inpr{U_F}{U_F}}$.
The gradients are then given by
\cite{PhysRevLett.120.150401}
\begin{align}
\partial_\alpha F(\Omega)
= -\mathrm{Re}
\left[ 
\frac{f^*(\Omega)}{\vert f(\Omega)\vert}
\frac{
\inpr{U_F}{
\partial_\alpha
U_\Omega}
}{\inpr{U_F}{U_F}}
\right].
\end{align}
While there is no analytical solution to the above equation, one can use the equation of motion
$\partial_t U_\Omega(t)
= -i H(t) U_\Omega(t)$
to obtain the coupled system of equations
\begin{align}
\partial_t
\left(
\begin{array}{c}
U_\Omega(t) \\
\partial_\alpha U_\Omega(t)
\end{array}
\right)
= -i 
\left(
\begin{array}{cc}
H & 0 \\
\partial_\alpha H & H
\end{array}
\right)
\left(
\begin{array}{c}
U_\Omega(t) \\
\partial_\alpha U_\Omega(t)
\end{array}
\right),
\end{align}
which can be solved by numerical forward integration, such as the adaptive Runge-Kutta. Machnes et al \cite{PhysRevLett.120.150401} demonstrated the GOAT algorithm by generating the iSWAP gate for a pair of transmon qubits coupled to a tunable bus resonator. Kirchoff et al demonstrated a time-optimal CNOT gate on a pair of transmon qubits \cite{kirchhoff2018optimized}.

\subsection{Variational methods \label{sec:varprinc}}
Variational methods are widely used for constrained optimization and are less sensitive to local optima compared to gradient methods.  For example, the Krotov algorithm that is based on the Lagrange multiplier exhibits a monotonic convergent behavior \cite{krotov1995global}.  It starts with the Lagrange multiplier 
\begin{align}
B_n & = 
\begin{cases}
U_{n+1:N}^\dagger U_F\inpr{U_F}{U_{1:N}} ~ &~\mbox{for gate synthesis, and}
\\
U_{n+1:N}^\dagger
\left\{\rho_F U_{1:N}\rho_I + \kappa U_{1:N}\right\}
~&~\mbox{for state-to-state transfer,}
\end{cases}
\end{align}
where $\kappa$ is a positive constant to ensure positivity of fidelity.  In addition to the control sequence $\Omega$, the Krotov method involves a co-sequence $\widetilde{\Omega}$, both of which are initialized with the same random guess, i.e.,
$\Omega^{(0)} = \widetilde{\Omega}^{(0)}$.
These sequences are propagated forward and backward respectively according to the rules \cite{doi:10.1063/1.2903458,doi:10.1063/1.3691827,batra2020push}
\begin{align}
	\omega_{mn}^{(i)} &= (1-\delta) \widetilde{\omega}_{mn}^{(i-1)} + \frac{\delta}{\lambda_m} \mathrm{Im} \inpr{B_n^{(i-1)}}{{\cal A}_m U_{0:n-1}^{(i)}}
	\nonumber \\
	\widetilde{\omega}_{mn}^{(i)} &= (1-\eta) \omega_{mn}^{{(i)}} + \frac{\eta}{\lambda_m} \mathrm{Im} \inpr{B_n^{(i)}}{{\cal A}_m U_{0:n}^{(i)}},
\end{align}
where $\lambda_m$ are penalty constants and other constants $\eta,\delta\in[0,2]$.
Application of Krotov optimization for the optimal control of spin dynamics in NMR and dynamical nuclear polarization gained significant attention \cite{doi:10.1063/1.2903458,doi:10.1063/1.3691827}. Vinding et al. showed the use of Krotov in magnetic resonance imaging for design of spatial selective RF pulses \cite{vinding2012fast}. Hwang et al. showed the control of a non-Markovian open quantum system for Z gate and identity gate \cite{PhysRevA.85.032321}. Krotov optimization has also been applied for the control of BEC in magnetic microtraps \cite{PhysRevA.90.033628}. 

\subsection{Machine Learning Methods
\label{sec:mlmtd}}
With the ever increasing computational power, the machine learning (ML) methods are finding applications everywhere, including QOC.
Reinforcement learning (RL), for example, is popularly used in a variety of problems of finding the best strategy that maximizes a reward function \cite{sutton2018reinforcement}. In QOC, RL can be modeled to find the strategy that maximizes the fidelity reward.  For example, Bukov et al demonstrated RL to control out-of-equilibrium systems \cite{PhysRevX.8.031086}.  RL has also been applied for problems such as state control \cite{An_2019}, gate control \cite{zhang2019does,baum2021experimental},  for generating controls robust against certain types of errors \cite{niu2019universal}, and for the control of multilevel dissipative quantum systems \cite{PhysRevA.103.012404}.
QOC with supervised ML \cite{PhysRevA.99.042327,ZENG2020126886} and  with convolutional neural networks trained through deep learning architecture for a quantum particle in a disordered system \cite{PhysRevApplied.17.024040} have also been reported.  Another type of ML, namely differential programming (DP), together with a neural network was used  for eigenstate preparation in a variety of single and multi-qubit systems  \cite{Sch_fer_2020} as well as for the control of quantum thermal machines \cite{khait2021optimal}. Coopmans et al. showed the transport of magnons in a spin chain by combining DP with CRAB and shortcut to adiabaticity protocols \cite{coopmans2022optimal}.

\subsection{Hybrid Methods
	\label{sec:hybridmtd}}
There have been several attempts to combine different algorithms to realize hybrid algorithms which incorporate the best of the original methods.
For example, Machnes et al. \cite{machnes2011comparing} combined GRAPE and Krotov to form the hybrid algorithm and implemented it in the DYNAMO package. 

While GRAPE can get stuck in a local optimum, the simulated  annealing (SA) is good at jumping over it and finding a better neighborhood. 
Thus the hybrid algorithm, SAGRAPE, combines the best of the two \cite{ram2021robust}. 
As explained in Sec. \ref{sec:simuanneal}, an important step in SA in every iteration is to scan the neighborhood points of the current solution, which is a bottleneck in the standard SAGRAPE algorithm.  This is where ML can bring about a significant speedup.  Priya et al \cite{batra2022recommender} have reported the recommender system (RS) expedited SAGRAPE (RSSAGRAPE), wherein both SA and GRAPE routines of SAGRAPE are speeded-up by RS.

The variation principle-based Krotov algorithm can lead to monotonic increase  at low fidelity but gets slow at high fidelities. The gradient-based quasi-newton BFGS can yield high fidelity solutions, but need more computational resources. To this end, Tannor et.al  \cite{eitan2011optimal} proposed the hybrid Krotov-BFGS (K-BFGS) algorithm.

For many QOC problems, high fidelity solutions lie in a low dimensional subspace of the larger dimensional search space \cite{lloyd2014information}. This property is exploited in algorithms like CRAB where the control sequence  is written as a linear combination of sensibly chosen basis functions which are fewer in number when compared  to the dimensionality of the original search space. 
Sorensen et.al \cite{sorensen2018quantum} proposed a hybrid algorithm called gradient optimisation using parameterisation (GROUP) that incorporates both CRAB and GRAPE algorithms. Here, an analytical expression of the gradient of the cost function w.r.t the previously mentioned linear coefficients is derived and is used in place of Nelder-Mead to update the linear coefficients. The resulting algorithm can produce higher fidelity controls for the same number of functional evaluations as the CRAB algorithm.    

\section{Summary and outlook \label{sec:summary}}
To summarize, we had set out three intentions in this review. First, to introduce the basic aspects of QOC.  Second, to point out various practical aspects in optimizing QOC as well as implementing the control sequence.  Third, to provide a quick overview of the diverse methods used for QOC.  We provided brief descriptions of various QOC algorithms sorting them into five categories, namely, asymptotic method, direct search, gradient method, variational method, and machine learning. 
A quick comparison of these methods is shown in Table. \ref{tab:comp}.  Finally, we also described hybrid algorithms which combine two or more methods to yield better performance.  

\begin{center}
\begin{table}[h]
\centering
\caption{An overview of the various QOC methods  discussed in the review.}
\label{tab:comp}    \begin{tabular}{p{2.5cm}p{2.5cm}p{4cm} p{6cm}}
\hline
Method & Algorithm & Merits & Limitations 
    \\ \hline   \hline
\multirow{3}[10]{4em}{Asymptotic evolution} & Adiabatic Pulses & Analytical control; Robust against pulse errors &  
Specialized applications; Lengthy sequences 
\\
   \cline{2-4}
   & Geometric Phase & Analytical  control;  Robust against pulse errors & Specialized applications; Lengthy sequences
       \\
    \cline{2-4}
    & Lyapunov & Analytical  control & Specialized applications
    \\ \hline    \multirow{3}[10]{4em}{Direct Search} & SMP & Highly adaptable & Sensitive to initial guess; Computationally inefficient
    \\ \cline{2-4}
    & Simulated Annealing & Can overcome local optima & Relatively slow
    \\ \cline{2-4}
    & CRAB & Analytical control sequence  & Sensitive to chosen model
    \\ \cline{2-4}
    & Evolutionary Algorithms & Highly adaptable; attempts global search  & Needs heavy computational resource
    \\ \hline
    \multirow{2}[3]{4em}{Gradient Method} & GRAPE  & Numerically efficient & Sensitive to initial guess 
    \\
    \cline{2-4}
    & GOAT & Analytical control sequence & Sensitive to chosen model
    \\ \hline
    Variational & Krotov  & Monotonic convergence & Needs higher computational resource
    \\ \hline 
    Machine Learning & Reinforcement Learning & Robust against local optima & A general ML platform for QOC needs to be developed
    \\ \hline
    Hybrid & & Can combine the best of two or more methods & Demand complex coding and fine-tuning of parameters
    \\ \hline
    \end{tabular}
\end{table}
\end{center}

Taking an outlook, we may expect further roles of hybrid algorithms as well as machine learning in designing complex and robust control sequences for larger systems with severe parameter constraints.  Measurement based feedback control (eg. \cite{lu2017enhancing}) which allows tailor-made controls incorporating device characteristics is likely to become routine.  Finally, a somewhat futuristic, but promising development is the quantum assisted quantum control, wherein quantum computers themselves are used to generate quantum controls (eg. \cite{PhysRevLett.127.220504}).

\subsubsection*{Software packages
	\label{sec:softpack}}
\textit{Simulation Package for Solid-state NMR spectroscopy} (SIMPSON), an open source package, simulates NMR experiments along with optimal control protocols such as GRAPE \cite{tovsner2009optimal}. SIMPSON package is available for different OS at \url{https://inano.au.dk/about/research-centers-and-projects/nmr/software/simpson}. \textit{Dynamic Optimization platform} (DYNAMO), a Matlab based software package contains GRAPE and Krotov algorithms along with the option of comparing and benchmarking new algorithms \cite{PhysRevA.84.022305}. The codes for the same can be found at \url{http://qlib.info}.  The \textit{Quantum Toolbox in Python} (QuTiP), an open source python package available at \url{https://qutip.org/download.html}, offers a simulation of generic quantum system with optimization techniques such as GRAPE and CRAB \cite{johansson2012qutip,JOHANSSON20131234}. Goerz et al. have implemented the Krotov algorithm in the QuTiP framework  \cite{10.21468/SciPostPhys.7.6.080}. Recently, Teske et al. developed an open source package QOPT for simulating quantum dynamics and robust quantum control in conjunction with common experimental situations \cite{PhysRevApplied.17.034036}. QEngine is a C++ library for the implementation of quantum control in ultra cold atoms \cite{SORENSEN2019135}. 

\section*{Acknowledgments}
The authors dedicate this review to the 80th birth day of Prof. Anil Kumar, IISc, Bangalore, who is noted for his pioneering contributions to NMR spectroscopy as well as NMR quantum computation.
PB acknowledges support from the Prime Minister’s Research Fellowship (PMRF) of the Government of India. TSM acknowledges funding from
DST/ICPS/QuST/2019/Q67.

\bibliography{revqctrl}

\end{document}